\documentclass[preprint,aps]{revtex4}
\makeatletter

\newcommand{\Rmnum}[1]{\expandafter\@slowromancap\romannumeral #1@}
\makeatother

\usepackage{graphicx}

\begin{document}
\title{Electronic Evidence for Type II Weyl Semimetal State in MoTe$_2$}
\author{Aiji Liang$^{1}$, Jianwei Huang$^{1}$, Simin Nie$^{1}$, Ying Ding$^{1}$, Qiang Gao$^{1}$, Cheng Hu$^{1}$, Shaolong He$^{1}$, Yuxiao Zhang$^{1}$, Chenlu Wang$^{1}$, Bing Shen$^{1}$, Jing Liu$^{1}$, Ping Ai$^{1}$, Li Yu$^{1}$, Xuan Sun$^{1}$, Wenjuan Zhao$^{1}$, Shoupeng Lv$^{1}$, Defa Liu$^{1}$, Cong Li$^{1}$, Yan Zhang$^{1}$, Yong Hu$^{1}$, Yu Xu$^{1}$, Lin Zhao$^{1}$, Guodong Liu$^{1}$, Zhiqiang Mao$^{2}$, Xiaowen Jia$^{3}$, Fengfeng Zhang$^{4}$, Shenjin Zhang$^{4}$, Feng Yang$^{4}$, Zhimin Wang$^{4}$, Qinjun Peng$^{4}$, Hongming Weng$^{1}$, Xi Dai$^{1,5}$, Zhong Fang$^{1,5}$, Zuyan Xu$^{4}$, Chuangtian Chen$^{4}$ and X. J. Zhou$^{1,5,*}$}
\affiliation{
\\$^{1}$Beijing National Laboratory for Condensed Matter Physics, Institute of Physics, Chinese Academy of Sciences, Beijing 100190, China,
\\$^{2}$Department of Physics and Engineering Physics, Tulane University, New Orleans, LA 70018
\\$^{3}$Military Transportation University, Tianjin 300161, China.
\\$^{4}$Technical Institute of Physics and Chemistry, Chinese Academy of Sciences, Beijing 100190, China.
\\$^{5}$Collaborative Innovation Center of Quantum Matter, Beijing, China,
\\$^{*}$Corresponding author: XJZhou@aphy.iphy.ac.cn.
}

\date{April 6, 2016}



\maketitle

{\bf Topological quantum materials, including topological insulators and superconductors, Dirac semimetals and Weyl semimetals, have attracted much attention recently for their unique electronic structure, spin texture and physical properties. Very lately, a new type of Weyl semimetals has been proposed where the Weyl Fermions emerge at the boundary between electron and hole pockets in a new phase of matter, which is distinct from the standard type I Weyl semimetals with a point-like Fermi surface. The Weyl cone in this type II semimetals is strongly tilted and the related Fermi surface undergos a Lifshitz transition, giving rise to a new kind of chiral anomaly and other new physics.  MoTe$_2$ is proposed to be a candidate of a type II Weyl semimetal; the sensitivity of its topological state to lattice constants and correlation also makes it an ideal platform to explore possible topological phase transitions. By performing laser-based angle-resolved photoemission (ARPES) measurements on MoTe$_2$ with unprecedentedly high resolution, we have uncovered electronic evidence of type II semimetal state in MoTe$_2$. We have established a full picture of the bulk electronic states and surface state for MoTe$_2$ that are consistent with the band structure calculations. A single branch of surface state is identified that connects bulk hole pockets and bulk electron pockets. Detailed temperature-dependent ARPES measurements show high intensity spot-like features that is $\sim$40 meV above the Fermi level and is close to the momentum space consistent with the theoretical expectation of the type II Weyl points. Our results constitute electronic evidence on the nature of the Weyl semimetal state that favors the presence of two sets of type II Weyl points in MoTe$_2$.}

The grafting of high energy physics concepts into the condensed matter physics with particles appearing in the form of low energy quasiparticle excitations has ignited the discovery of numerous topological quantum materials which are represented by topological insulators and superconductors\cite{X. L. Qi,M. Z. Hasan}, Dirac semimetals\cite{Z. J. Wang1,Z. K. Liu1,S. Y. Xu1,Z. J. Wang2,Z. K. Liu2,M. Neupane,S. Borisenko,H. M. Yi}, Weyl semimetals\cite{H. Weyl,H. B. Nielsen1,S. Murakami1,X. G. Wan,A. A. Burkov,G. Xu,S. M. Huang,H. M. Weng,S.Y.Xu2,B. Q. Lv,L. X. Yang} and so on. Very lately, it was shown that Weyl Fermions in lattice systems can be further classed into two types\cite{A. Soluyanov}. In type I Weyl semimetals, a topologically protected linear crossing of two bands, called a Weyl point, occurs at the Fermi level resulting in a point-like Fermi surface\cite{A. Soluyanov}. On the other hand, when  the Lorentz invariance is violated in condensed matter, a fundamentally new kind of Weyl Fermions can be produced. In the type II Weyl semimetals, the Weyl point emerges from a contact of an electron and a hole pocket at the boundary resulting in a highly tilted Weyl cone. The emergence of this type II Weyl Fermions leads to new physical properties that are very different from those of the standard Weyl semimetals\cite{A. Soluyanov}.

The proposed material candidates for realizing type II Weyl semimetals include WTe$_2$, MoTe$_2$ and their mixture (W,Mo)Te$_2$\cite{A. Soluyanov,Y. Sun,T. R. Chang,Z. J. Wang3,I. Belopolski}.  For MoTe$_2$ crystalized in Td form which is in orthorhombic phase with no inversion symmetry (Fig. 1a) and isostructural to WTe$_2$, it has been expected to have a larger separation of Weyl points in momentum space and more extended arc surface states, more desirable for unveiling Weyl points by spectroscopic tools like angle-resolved photoemission spectroscopy (ARPES)\cite{Y. Sun,Z. J. Wang3}. It has been found by band structure calculations that the topological nature associated with the type II Weyl physics in MoTe$_2$ is extremely sensitive to its lattice constants. Small variations of lattice constants result in different forms of Weyl semimetals ranging from showing 4 sets of type II Weyl points\cite{Y. Sun,I. Belopolski},  to 2 sets of type II Weyl points\cite{Z. J. Wang3}, to 2 sets of type I Weyl points\cite{Y. Sun}. Therefore, MoTe$_2$ is interesting not only because of its being a viable candidate for a type II Weyl semimetal, but also its possibility of topological phase transitions realized through delicate tuning of strain by pressure or temperature\cite{Y. Sun,I. Belopolski,Z. J. Wang3}.

In this work, we report our high resolution laser-based ARPES results on identification of the electronic signatures of the type II Weyl semimetals  as well as the possible temperature-induced topological phase transition in MoTe$_2$. Through our comprehensive ARPES measurements with unprecedentedly high resolution, we have established a full picture of the bulk electronic structures and surface state in MoTe$_2$ that are consistent with band structure calculations. A single branch of surface state is identified that connect the bulk hole pockets and the bulk electron pockets.  Detailed temperature-dependent measurements reveal high intensity spot-like features that is $\sim$40 meV above the Fermi level in the momentum space that is in agreement with the expected type II Weyl points from theoretical calculations.  The electronic structure of MoTe$_2$ is robust against temperature variation and no signature of a topological phase transition is seen in the temperature range we have covered.  Our results constitute evidence on the observation of type II  semimetal states in MoTe$_2$ that favors the existence of two sets of type II Weyl points. 

Figure 1 shows the detailed Fermi surface and band structures of MoTe$_2$ and their comparison with band structure calculations. Here we employed two latest generation laser-based ARPES systems with super-high energy and momentum resultions to do the measurements (see Supplementary for sample and experimental details). The ARToF-ARPES system can cover two-dimensional momentum space simultaneously; this is advantageous in resolving multiple fine electronic details because of its very dense momentum coverage.  Fig. 1c and 1d show Fermi surface of MoTe$_2$ measured at 25 K under two different polarization geometries by using this ARToF-ARPES system. The other DA30-ARPES system can provide fine automatic Fermi surface mapping along multiple momentum cuts without rotating the sample; it is suitable for taking high resolution band structure measurements, as shown in Fig. 1i measured on MoTe$_2$ along some representative momentum cuts indicated in Fig. 1h. Because of the matrix element effects involved in the photoemission process, the measured photoelectron intensity varies sensitively with the measurement condition like photon energy, photon polarization and sample orientation, as seen from different band intensities in Fig. 1c and 1d when {\it s}- and {\it p}-polarization geometries  are employed. Our ARPES measurements under these two distinct polarization geometries (Fig. 1c and 1d) can provide complementary information to fully reveal the electronic structure for the material measured.

Careful examination of the Fermi surface shown in Fig. 1c and 1d, combined with the band structure analysis (Fig. 1i), allows us to get a full picture of the Fermi surface, as schematically shown in Fig. 1e, and the band structures in Fig. 1i. Further comparison with the calculated bulk Fermi surface (Fig. 1f) and calculated Fermi surface including the surface state (Fig. 1g), makes it possible to pin down on the nature of each Fermi surface sheet and band. We can identify three bulk hole pockets and two bulk electron pockets that are symmetrically distributed along $\bar{\Gamma}$-$\bar{X}$ direction with respect to the Brillouin zone center $\bar{\Gamma}$.  The three hole pockets are denoted by H1 for the circular hole pocket around the Brillouin zone center, and H3 and H4 for the two butterfly-like hole pockets (Fig. 1e). The two electron pockets are denoted as E1 and E2 (Fig. 1e). These bulk electronic structures  show a good agreement with band structure calculations (Fig. 1f) in terms of the number of hole- and electron-like Fermi surface sheets, and their locations in the momentum space. On the other hand, there are deviations between the measurements and calculations regarding the exact shape and size of each Fermi surface sheet that may come from the effect of k$_z$, hole doping effect, and/or relative shifting of bands.

The band structures depend sensitively on k$_z$ for a three-dimensional material like MoTe$_2$. In particular, k$_z$ is important for probing the Weyl Fermion physics in MoTe$_2$ because it is expected that Weyl Fermions emerge only for k$_z$=0\cite{A. Soluyanov,Y. Sun,T. R. Chang,Z. J. Wang3,I. Belopolski}. We have performed detailed comparison of the measured band structures with the calculated ones at different k$_z$s (Fig. 2). Focusing on some characteristic bands like $\delta$ and $\varepsilon$ bands that vary their energy positions sensitively with k$_z$ (Fig. 2a and 2b, and also see Supplementary Materials), we conclude that, for the photon energy of 6.994 eV we used in this work, the corresponding k$_z$ is close to 0.

Our measurements have clearly identified  surface states in MoTe$_2$. As seen in Fig. 1c and 1d, and also schematically shown in Fig. 1e, there exists a segment labeled by SS that is between the H4 hole pocket and the E1 electron pocket.  This SS segment does not appear in the bulk band structure calculations (Fig. 1f), but only emerges in the band structure calculations including surface states (Fig. 1g).  The measured band structure of this particular ss band (Fig. 2c) shows a clear resemblance to the surface state band in the band structure calculations (Fig. 2d). The constant energy contours of this SS segment at different binding energies (Fig. 2e and 2f) also show good agreement with those of the surface band from the band structure calculations (Fig. 2g). These observations provide strong evidence on the surface state nature of the SS band in MoTe$_2$.

Our observation of the hole pocket H4, electron pocket E2 and the surface state segment SS (Fig. 1e) makes it possible to investigate their relationship that is important in determining the nature of the surface state and the Weyl physics. To this end, we carried out ARPES measurements with enhanced high-momentum resolution, focusing on the momentum region of possible Weyl points measured under two distinct polarization geometries, as shown in Fig. 3(a-d). Surface state dominates in the {\it s}-polarization measurements (Fig. 3a and 3b) while bulk bands are enhanced in the {\it p}-polarization geometry (Fig. 3c and 3d), but the electronic structures measured under these two polarization geometries are similar. They are complementary to get full information on the related bands. Following results can be obtained from Fig. 3. (1). We can simultaneously resolve hole band h4, electron band e2 and the surface state band ss clearly in Fig. 3.  Along or close to the $\bar{\Gamma}$-$\bar{X}$ cut, the surface state band ss and the hole band h are separated at the Fermi level, but they touch at a binding energy of 80 meV forming a tilted  X-like shape (cuts 1 to 3 in Fig. 3d).  When the momentum cuts move away from the $\bar{\Gamma}$-$\bar{X}$ line (cuts 4 to 7 in Fig. 3d), the two bands get close to each other and merge together at k$_y$ ranging from $\sim$0.04 {\AA}$^-$$^1$ to  $\sim$0.08 {\AA}$^-$$^1$, but they separate at high binding energies.  When the momentum cuts goes further away from $\bar{\Gamma}$-$\bar{X}$ line, the surface state ss and the hole band h gradually separate from each other again. The h  band split into h3  and h4 bands to generate butterfly-like hole pockets H3 and H4. These give rise to an overlap area of the surface state segment SS and the hole pocket H4, as shown in Fig. 1e. (2). With the electron band e2 and hole band h4 resolved clearly, upon careful examination, we observe only a single branch of surface state ss in the whole k$_y$ range of 0 to 0.1 {\AA}$^-$$^1$. (3). As seen from Fig. 1e and Fig. 3, the surface state ss comes out from the bulk electron band(s). As seen in Fig. 3, the surface state ss overlaps with the bulk hole band h4 near k$_y$=0.05 {\AA}$^-$$^1$. Therefore, the surface state SS connects the bulk electron pocket and the hole pocket as shown in Fig. 1e. (4). In the momentum space that we have covered, the electron band e2 is always separated from the surface state band ss at the Fermi level. However, these two bands are close to each other. Moreover, the electron band e2 is rather flat while the surface state band and hole band are steep. From temperature-dependent band structure measurements (Fig. 3(e-k)), by empirical extrapolation, we expect that the electron band e2 and the hole band h may touch each other at an energy of $\sim$40 meV above the Fermi level. The above results indicate the existence of the Weyl points in MoTe$_2$. 

The Weyl points in MoTe$_2$ have been expected to lie above the Fermi level in the unoccupied states\cite{A. Soluyanov,Y. Sun,I. Belopolski,Z. J. Wang3}.  In order to observe the Weyl points, we carried out detailed temperature-dependent ARPES measurements on MoTe$_2$ up to high temperatures (200 K). ARPES measurements at high temperature make it possible to observe electronic states above the Fermi level due to thermal excitations. Fig. 4(a-j) shows ARPES measurements at different temperatures using ARToF-ARPES under two distinct polarization geometries. These extensive measurements cover the momentum space and energy space above the Fermi level where possible Weyl points are expected\cite{A. Soluyanov,Y. Sun,I. Belopolski,Z. J. Wang3}.  In the {\it s}-polarization measurements, the surface state SS dominates the signal at the Fermi level (Fig. 4(a-e)). The surface state SS gets weaker when the energy goes above the Fermi level; in this case, it is possible to identify the electron pocket E2, particularly in Fig. 4b  for 20 meV energy at 60 K (marked by the black dashed line) and Fig. 4c at 100 K.  With increasing energy, the electron pocket E2 gradually approaches the surface state SS.  Such a behaviour is natural and consistent with the results in Fig. 3 because electron band E2 is rather flat while the surface state band ss is steep.  

On the other hand, under the {\it p}-polarization (Fig. 4(f-j)), at a low temperature (30 K), the most pronounced features at the Fermi level (Fig. 4f) are bulk bands: hole pocket H1 and electron pocket E1. The H3 and H4 hole pockets are weaker but still clear. The intensity of the surface state SS is much suppressed in the {\it p}-polarization (Fig. 4(f-j)) compared with that in the {\it s}-polarization (Fig. 4(a-e)).  When the energy goes above the Fermi level, the surface state SS is quickly suppressed.  A peculiar feature stands out, i.e.,  the strong intensity spots in the measured constant energy contours, as marked by the arrows in Fig. 4(h-j).  The strong intensity spot becomes clearer at $\sim$20 meV and above in the high temperature measurements. We note that, different from the {\it s}-polarization measurements where the left and right intensities of the surface state SS are nearly symmetric with respect to the $\bar{\Gamma}$-$\bar{X}$ line, the strong spots on the right side are much weaker in the {\it p}-polarization, likely due to photoemission matrix element effects.  With the overall intensity of the electron pocket E2, the hole pocket H3, and the surface state SS that are vanishingly weak at $\sim$40 meV, the appearance of the strong intensity spot indicates something robust and nontrivial. From Fig. 4(k-m), one can see that the spectral weight distribution evolves from a smooth distribution at the Fermi level to the emergence of a discrete strong intensity spot centered around (0.225,$\pm$0.06){\AA}$^-$$^1$ at an energy of $\sim$40 meV above the Fermi level (Fig. 4k). With the merging of the electron pocket E2 with the surface state SS segment above the Fermi level, as well as the contact of the SS surface state segment with H4 hole pocket, this strong spot is situated near the overlapping region of the electron pocket E2 and hole pocket H4.  The momentum location and the associated energy ($\sim$40 meV) of the strong spots are close to the Weyl points expected from theoretical calculations\cite{A. Soluyanov,Y. Sun,I. Belopolski,Z. J. Wang3}.

It has been shown by band structure calculations that, depending on the lattice constants and electron correlations, different nature of Weyl points may result, as schematically shown in Fig. 4(n,o,p)\cite{A. Soluyanov,Y. Sun,I. Belopolski,Z. J. Wang3}.  Our detailed temperature-dependent measurements in Fig. 4 also makes it possible to examine on the possible topological phase transition induced by the lattice constant change from the temperature variation. As seen in Fig. 4, for a given energy at the Fermi level or above the Fermi level, the observed electronic structures including both bulk bands and surface states are rather robust to the temperature variation.  This indicates that, in the temperature range we have covered (30-200 K), the strain from lattice change may not be large enough to induce the topological phase transition that is expected.


Our comprehensive ARPES data provides a clear distinction between different nature of the Weyl points in MoTe$_2$. Depending on the lattice constants and electron correlation, it has been proposed that MoTe$_2$ may have three kinds of Weyl points: (I). four sets of type II Weyl points (Fig. 4n)\cite{Y. Sun}; (II). two sets of type I Weyl points(Fig. 4o)\cite{Y. Sun}; and (III). two sets of type II Weyl points(Fig. 4p)\cite{Z. J. Wang3}. Our results are apparently not consistent with the scenario II of two sets of type I Weyl Fermions because the observed Weyl point is formed from the contact of the electron pocket E2 and the hole pocket H4, as seen from Fig. 3. When it comes to the other two kinds of type II Weyl Fermions, there is a decisive difference between them in the number of surface states between the Weyl points.  For the scenario I with four sets of type II Weyl points (Fig. 4n), two surface states are expected: one is trivial while the other is a Fermi arc connecting the two Weyl points\cite{Y. Sun}. For the scenario III with two sets of type II Weyl points, only one surface state is expected in the similar momentum region (Fig. 4o)\cite{Z. J. Wang3}. Our clear observation of only a single surface state in the momentum space of interest, in between the electron band e2 and the hole band h4 that are well resolved, favors scenario III, i.e., there exist two sets of type II Weyl points in MoTe$_2$.

In summary,  we have established a full picture of the bulk electronic states and surface state for MoTe$_2$ that are consistent with the band structure calculations. A single branch of surface state is identified that connects bulk hole pockets and bulk electron pockets. Detailed temperature-dependent ARPES measurements show high intensity spot-like features that is $\sim$40 meV above the Fermi level and is close to the momentum space consistent with the theoretical expectation of the type II Weyl points. Our results constitute electronic evidence on the nature of the Weyl semimetal state that favors the presence of two sets of type II Weyl points in MoTe$_2$.


\vspace{3mm}

\noindent {\bf Acknowledgement}\\
This work is supported by the National Science Foundation of China (11574367), the 973 project of the Ministry of Science and Technology of China (2013CB921700, 2013CB921904 and 2015CB921300) and the Strategic Priority Research Program (B) of the Chinese Academy of Sciences (Grant No. XDB07020300).

\vspace{3mm}

\noindent {\bf Author Contributions}\\
X.J.Z. and A.J.L. proposed and designed the research. Z.Q.M. contributed in sample growth. S.M.N., H.W.M., X.D. and Z.F. contributed in the band structure calculations. A.J.L., J.W.H., Y.D., Q.G., C.H., S.L.H., Y.X.Z., C.L.W., B.S., J.L., P.A., L.Y., X.S., W.J.Z., S.P.L., D.F.L., C.L., Y.Z., Y.H., Y.X., L.Z., G.D.L., X.W.J., F.F.Z., S.J.Z., F.Y., Z.M.W., Q.J.P., Z.Y.X., C.T.C. and X.J.Z. contributed to the development and maintenance of the Laser-ARTOF and Laser-DA30 ARPES systems and related software development. A.J.L., J.W.H., Y.D., Q.G. and C.H. carried out the ARPES experiment. A.J.L. and X.J.Z. analyzed the data. X.J.Z. and A.J.L. wrote the paper with S.L.H., S.M.N., G.D.L., J.W.H., C.L.W. and Q.G.. All authors participated in discussion and comment on the paper.\\

\noindent{\bf Additional information}\\
Supplementary information is available in the online version of the paper.
Correspondence and requests for materials should be addressed to X.J.Z.



\newpage

\begin{figure*}[tbp]
\centering
\includegraphics[width=1.0\textwidth]{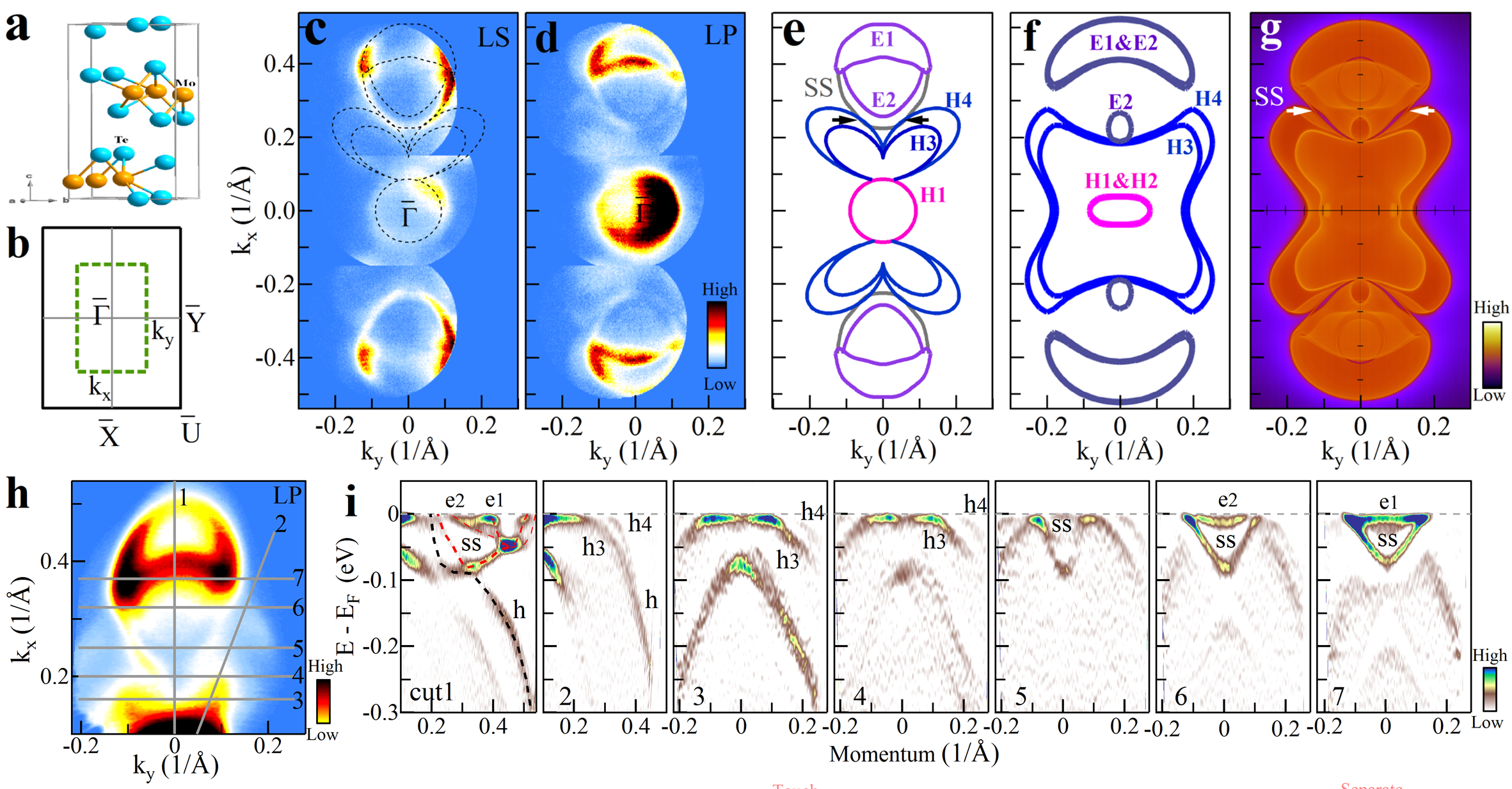}
\caption{\label{Figure 1}\textbf{Electronic structure of MoTe$_2$ and its comparison with calculated results.} (a) Crystal structure of MoTe$_2$ in Td form with a space group \emph{Pmn}2$_1$ and lattice constants of a=3.48{\AA}, b=6.33{\AA}, c=13.798{\AA}. (b) Projected surface Brillouin zone of the (001) surface. The green dashed rectangle marks the momentum region covered in our ARPES measurements. (c) Fermi surface of MoTe$_2$ taken by ARToF-ARPES at 25 K with {\it s}-polarization geometry. The spectral weight distribution is obtained by integrating photoemission spectra at each momentum within an energy window of [-2.5,2.5]meV with respect to the Fermi level. The black dashed lines are guides to the eye showing the measured Fermi surface contours.  (d) Fermi surface of Td-MoTe$_2$ taken by ARToF-ARPES at 25 K with {\it p}-polarization geometry.  In both (c) and (d), the lower 1/3 region is obtained by symmetrizing the upper 1/3 part of the measured Fermi surface with respect to the $\bar{\Gamma}$-$\bar{Y}$ mirror plane. (e) Schematic of all measured Fermi surface sheets and their assignments. H1, H3 and H4 represent measured bulk hole pockets, E1 and E2 represent measured bulk electron pockets, and  SS represents observed surface state segments. The touching of the surface state segments SS and the bulk hole pockets H4 are marked by black arrows. (f) Calculated bulk Fermi surface at k$_z$=0. (g) Calculated projected Fermi surface including surface states. The white arrows mark the surface state contours. (h) Fermi surface of Td-MoTe$_2$ measured by DA30-ARPES at 25 K with {\it p}-polarization geometry.  (i) Band structures along several typical momentum cuts. The location of the momentum cuts are marked by grey lines in (h). Here the band images are obtained by second derivative of the original data with respect to energy to highlight the band features.
}
\end{figure*}

\begin{figure*}[tbp]
\centering
\includegraphics[width=1.0\textwidth]{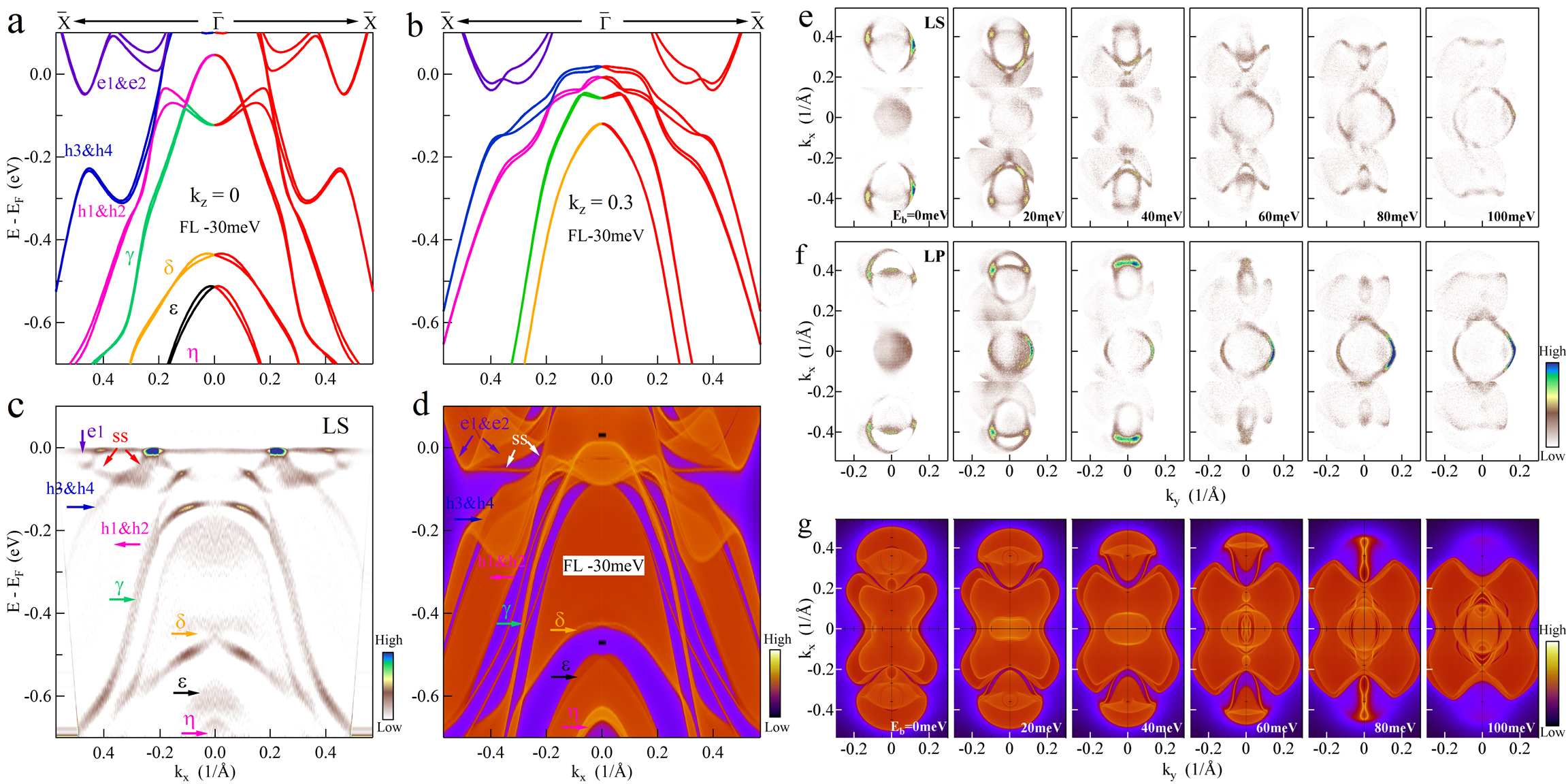}
\caption{\label{Figure 2}\textbf{Identification of surface state SS in MoTe$_2$.} (a) Calculated bulk band structure along $\bar{X}$-$\overline{\Gamma}$-$\bar{X}$ with k$_z$ fixed at 0 and the Fermi level is shifted downwards by 30 meV. Bands are labeled with different colours and letters. (b) Same as (a) but with k$_z$ at 0.3 2${\pi}$/c. (c) Band structure of MoTe$_2$ measured along $\bar{X}$-$\bar{\Gamma}$-$\bar{X}$ under {\it s}-polarization geometry at 30 K. The band images are obtained by second derivative of the original data with respect to energy. (d) Calculated band structure including the surface state. The Fermi level is shifted 30 meV downwards for comparison with the measured results in (c). (e) Constant energy contours of MoTe$_2$ at different binding energies measured at 30 K under {\it s}-polarization geometry.  (f) Constant energy contours of MoTe$_2$ at different binding energies measured at 30 K under {\it p}-polarization geometry. (g) Calculated constant energy contours including surface states at different binding energies .
}
\end{figure*}

\begin{figure*}[tbp]
\centering
\includegraphics[width=1.0\textwidth]{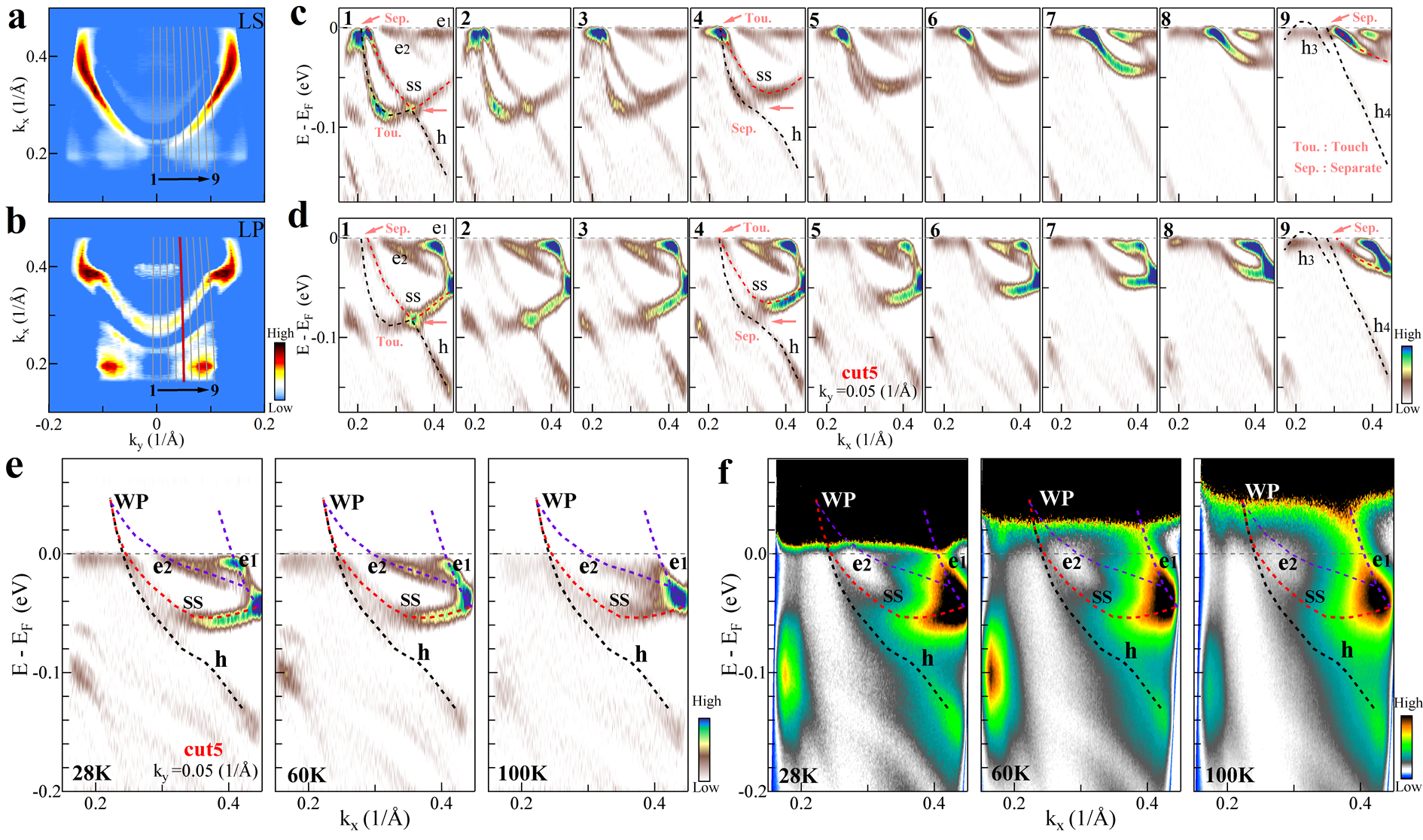}
\caption{\label{Figure 2}\textbf{Relation between the surface state and the bulk states in MoTe$_2$.} (a) Fermi surface of MoTe$_2$ 
measured by DA30-ARPES with enhanced high angular resolution at 25 K under {\it s}-polarization geometry. The image is obtained by the second derivative of original data with respect to energy and is symmetrized with respect to the $\bar{\Gamma}$-$\bar{X}$ mirror plane. (b) Same as (a) but measured under {\it p}-polarization geometry.  (c) Band structures measured along the momentum cuts marked in (a) as grey lines. The band images are obtained by second derivative of the original data with respect to energy to highlight the band features. They show clear momentum evolution of the surface state ss, the bulk hole band h and the bulk electron band e2. (d) Same as (c) but measured under {\it p}-polarization geometry. (e) Band structures measured at different temperatures along the momentum cut 5. Here the EDC second derivative images are shown. (f) Band structures measured at different temperatures along the momentum cut 5. Here the original data are shown but with the Fermi distribution functions removed in order to show states above the Fermi level.
}
\end{figure*}

\begin{figure*}[tbp]
\centering
\includegraphics[width=1.0\textwidth]{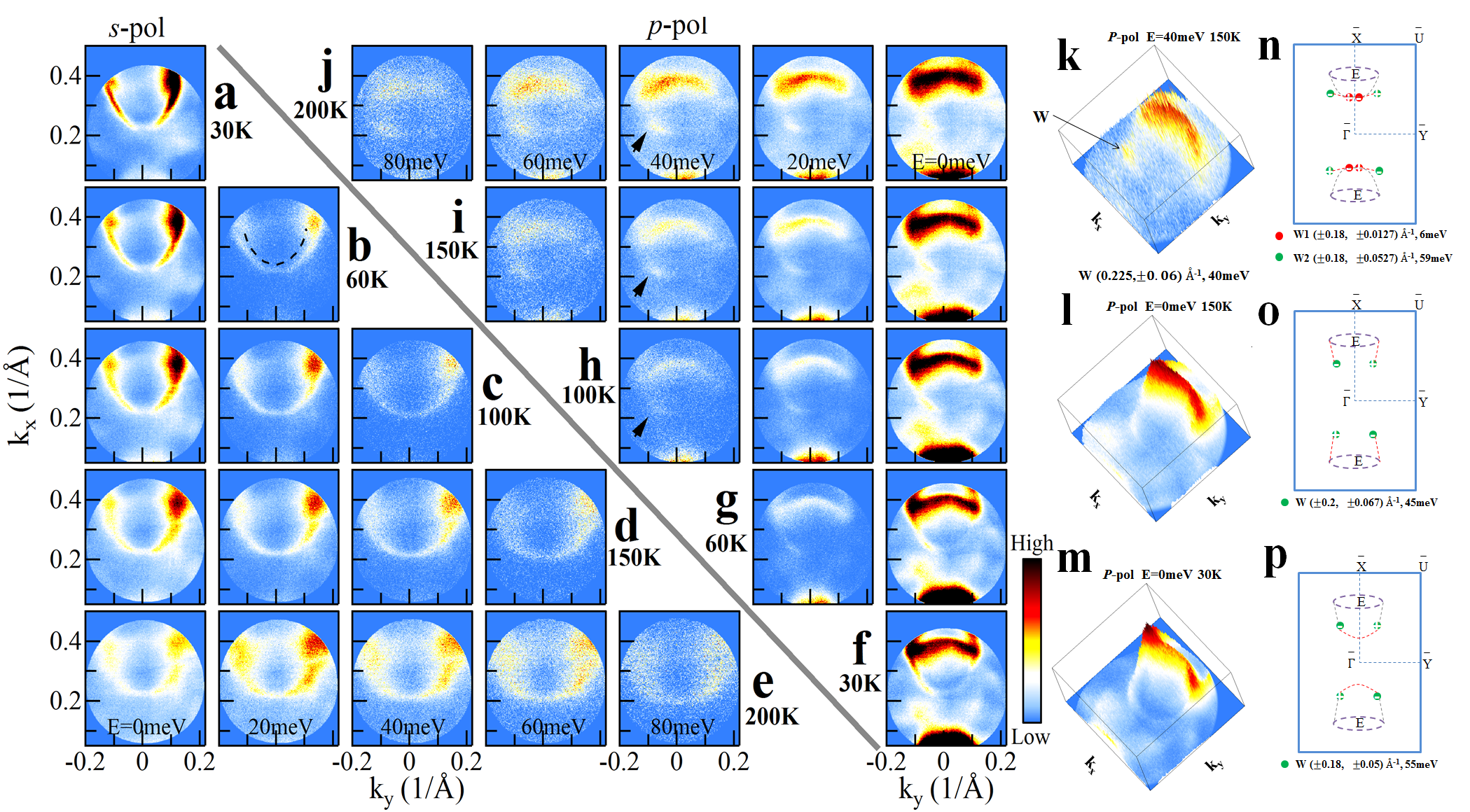}
\caption{\label{Figure 3} \textbf{Constant energy contours of MoTe$_2$ at different temperatures and different energies above the Fermi level.} (a)-(e) Constant energy contours at different energies above the Fermi level measured by ARToF-ARPES in {\it s}-polarization at different temperatures of 30K, 60K, 100K, 150K and 200K, respectively.  The constant energy contours are obtained by integrating the spectral weight within an energy window of $\pm$10 meV with respect to each indicated energy. Black dashed line in 20 meV, 60 K panel of (b) shows the trajectory of the weak E2 pocket. (f-j) Same as (a-e) but measured under {\it p}-polarization.  The high intensity spot marked by black arrows in the 40 meV constant energy contours at 100 K (h), 150 K (i) and 200 K(j) might point to the type II Weyl points that are consistent with predictions in Ref. \cite{Y. Sun,Z. J. Wang3}. (k-m) shows three-dimensional plots of the spectral intensity distribution as a function of momentum k$_x$ and k$_y$ measured under {\it p}-polarization geometry. The spectral weight shows a smooth variation at the Fermi level for 30 K (m) and 150 K (l), but a discrete strong intensity spot emerges near (0.225,$\pm$0.06){\AA}$^-$$^1$ when the energy is at 40 meV (k).  (n-p) shows three possible scenarios of Weyl Fermions in MoTe$_2$ with four sets of type II Weyl points (n)\cite{Y. Sun}, two sets of type I Weyl points (o)\cite{Y. Sun} and two sets of type II Weyl points (p)\cite{Z. J. Wang3}.
}
\end{figure*}

\end{document}